\documentclass[aip,graphicx]{revtex4-1}

\usepackage{latexsym}
\usepackage{graphics}
\usepackage{dsfont}
\usepackage{amsmath,amssymb,comment,float, subfig, adjustbox}
\usepackage{xcolor}

\draft % marks overfull lines with a black rule on the right

\begin{document}

%\preprint{}

\title{Modeling flywheel energy storage system charge and discharge dynamics} %Title of paper



\author{Pieter-Jan C. Stas}
%\thanks{}
\affiliation{Department of Applied Physics, Stanford University 348 Via Pueblo Mall, Stanford, California 94305, USA}

\author{Sulav Ghimire}
%\thanks{}
\affiliation{Center for Energy Science and Technology, Skolkovo Institute of Science and Technology, 3 Nobel Street, Skolkovo, Moscow Region 121205, Russia}

\author{Henni Ouerdane}
%\thanks{}
\affiliation{Center for Energy Science and Technology, Skolkovo Institute of Science and Technology, 3 Nobel Street, Skolkovo, Moscow Region 121205, Russia}


\date{\today}

\begin{abstract}
Energy storage technologies are of great practical importance in electrical grids where renewable energy sources are becoming a significant component in the energy generation mix. Here, we focus on some of the basic properties of flywheel energy storage systems, a technology that becomes competitive due to recent progress in material and electrical design. While the description of the rotation of rigid bodies about a fixed axis is classical mechanics textbook material, all its basic aspects pertinent to flywheels, like, e.g., the evaluation of stress caused by centripetal forces at high rotation speeds, are either not covered or scattered over different sources of information; so it is worthwhile to look closer at the specific mechanical problems for flywheels, and to derive and clearly analyse the equations and their solutions. The connection of flywheels to electrical systems impose particular boundary conditions due to the coupling of mechanical and electrical characteristics of the system. Our report thus deal with the mechanical design in terms of stresses in flywheels, particularly during acceleration and deceleration, considering both solid and hollow disks geometries, in light of which we give a detailed electrical design of the flywheel system considering the discharge-stage dynamics of the flywheel. We include a discussion on the applicability of this mathematical model of the electrical properties of the flywheel for actual settings. Finally, we briefly discuss the relative advantages of flywheels in electrical grids over other energy storage technologies. 
\end{abstract}


\maketitle %\maketitle must follow title, authors, abstract and \pacs

\section{Introduction}
     Energy storage technologies are key components of modern energy systems \cite{Bradford2018}, which are currently undergoing a transition to significantly reduce fossil fuel consumption \cite{Ongena2018}; these technologies are varied since energy can be harvested, converted and stored under different forms, broadly speaking: mechanical, chemical, thermal, and electrical \cite{Akinyele2014}. Of course, these technologies are not equivalent in terms of usage, size, storage capacity, conversion efficiency, ability to release power fast when needed, and lifecycle management; for instance a hydroelectric dam (1000 MW capacity) serves a purpose different than that of a battery for a cell phone (1000 mW) and costs and management are also on largely different scales. More specifically for electrical grids and power stations, energy storage solutions include, e.g., compressed air, pumped water, flywheels, batteries, and hot water tanks to name a few \cite{Akinyele2014,Wagner}. A solution can also be based on the combination of two technologies, e.g., batteries and flywheels; these latter which offer the advantage of being able to store large quantities of mechanical energy that can be delivered at high electrical power, may offset the detrimental ageing effects of battery charging and discharging cycles by saving battery power. 
     
     With the growing penetration of renewable but intermittent energies like solar and wind in the electrical grids, the problem of intermittency management while ensuring balance between the supply and demand of electricity generation, becomes increasingly challenging \cite{Reihani2016,Hammons2008}. Matching the energy demand that fluctuate over different time scales with an energy supply that has a growing unpredictable character, requires widespread energy storage technologies that facilitate demand response \cite{Wagner,romanelli}. Energy storage systems specialized to electrical grids store large amounts of energy when surplus electricity is produced, to later supply it when demand increases, but these crucial power system components should ideally also conform to various constraints including, e.g., fast response; capex and opex; lifecyle; reliability; sustainability; conversion efficiency. As mentioned above, solutions are varied, and choices must account for the constraints. Among the different technologies, flywheels, which are the object of the present work, attract growing attention owing to progress in materials science, mechanical bearing design, and connection to power electronics equipment \cite{Amiryar2017,sonsky}. 
    
    %\begin{figure}
    %\centering
    %\resizebox{0.7\textwidth}{!}{\includegraphics{peak_shaving_no_units.png}}
    %\caption{Qualitative representation of a typical load behaviour, and the role that energy storage plays in flattening the load. In the middle of the day, the surplus of energy is stored in a). Later in the day, this stored energy is used in b) to balance the underproduction.}
    %\label{fig:peakshave}
    %\end{figure}
    
    A flywheel is a solid mass that can rotate freely around a given axis, and hence can store electrical energy under a mechanical (rotational kinetic) form. The flywheel is by no means a novel invention; in fact, flywheel mechanisms have been used by humans for millennia, such as in millstones, hand looms, and potters wheels. With the industrial revolution, flywheels started being used as a short-term energy storage solution to stabilize the power output of steam engines by smoothing energy pulses of engine pistons somehow in a fashion similar to smoothing capacitors in electric circuits \cite{breeze}. In fact, the use of flywheels has typically been limited to low-energy applications for essentially two reasons: i/ the ``traditional'' materials these are made of: iron or lead, which make high-density energy storage difficult to attain, as at high speeds, the damaging effects (stress) of centripetal forces become harder and harder to withstand; ii/ the mechanical bearings generate friction that results in the rapid loss of stored energy. Two technological advances now make flywheel energy storage systems (FESS) a promising long-term energy storage option. First, experimentation with materials such as epoxy, kevlar, and carbon compound flywheels have pushed the limits of high-energy density, high-rotation speed flywheels \cite{kale} (see Table~\ref{matprop} for figures); next, replacing of mechanical bearings with (electro)magnetic bearings and placing the flywheel in vacuum chambers allow for FESS with minimal loss rates and the possibility to use them for long-term storage \cite{sonsky,nasa}.
    
    One important quantity that characterizes the usefulness of flywheels is the specific energy, or energy density, which is a metric that allows direct comparison with other current energy storage technologies, such as batteries or supercapacitors. The specific energy of flywheels, $e$, is given by \cite{genta}:
    
    \begin{eqnarray}
    e = K \frac{\sigma_u}{\rho}
    \label{eq:spece}
    \end{eqnarray}
    
    \noindent where $\sigma_u$ is the ultimate strength, $\rho$ is the density of the material, and $K$ is the shape factor that depends on the geometry and properties of the material. The capacity to withstand stress at high rotation speeds, and how the capacity changes under acceleration or deceleration, with the geometry, and under different temperature and humidity conditions, are  critical flywheel characteristics that may be studied experimentally with burst containment test rigs \cite{Buchroithner2018}, and also analytically or numerically. The basic description of the motion of a flywheel rests on standard textbook mechanics \cite{Taylor2004}, and nothing fundamentally new can be expected on this front; but accurate calculations of the stress tensor specialized to flywheels and their analysis, are often missing. Further, the detailed analysis of the flywheel electromechanical properties also lacks in the context of energy storage pertinent to electrical grids. In this work, we aim to contribute a didactic, integrated, and self-contained overview of FESS design. We particularly give attention to the mechanical design and the behavior of different stress components both for constant angular velocity rotation and during acceleration and deceleration, also accounting for the coupling to the electrical system. We aim to keep our calculations on the analytical level to provide results, which are physically transparent, and which can serve also as limiting cases for the test of numerical simulations of complex FESS structures. 
    
    \begin{table}
    \centering
    \caption{Ultimate tensile strength, volumetric density for different materials \cite{peirson,etb}. Poisson ratios, $\nu$, are also given. Traditional materials are shown separately from the newer composites. The values provided for the specific energy $e$, are calculated with $K=1$.}
    \label{matprop}
    \begin{tabular}{lllll}
    \hline
    \hline\noalign{\smallskip}
    Material & $\sigma_u$ [MPa] & $\rho$ [kg$\cdot$m$^{-3}$] & $e$ [kJ$\cdot$kg$^{-1}$] & $\nu$\\
    \noalign{\smallskip}\hline\noalign{\smallskip}
    Iron & 900 & 8000 & 112.5 & 0.22 - 0.30\\
    Titanium alloy & 650 & 4500 & 144.4 & 0.26 - 0.34\\
    Brass & 330 & 8410 & 39.2 & 0.33 - 0.36\\
    \noalign{\smallskip}\hline\noalign{\smallskip}
    CC fibre (40$\%$ epoxy) & 750 & 1550 & 483.9 & 0.25\\
    Kevlar fibre (40$\%$ epoxy) & 1000 & 1400 & 714.3 & 0.36\\
    \noalign{\smallskip}\hline\hline
    \end{tabular}
    \end{table}
    
    The article is organized as follows. In section 2 we present an analysis of the mechanical properties of flywheels. In section 3 we turn to the electrical design, for the charging, storage, and discharging processes, as well as the stabilization with magnetic bearings. In section 4 we discuss the relative advantages of flywheels in electrical grids in comparison to other energy storage technologies.

\section{Mechanical design of flywheels}

\subsection{Constant angular velocity rotation}

    Assuming that the flywheel experiences no external force along its rotation axis ($z$-axis), the  mathematical description of its dynamics reduces to a 2-dimensional problem with axes $r$ (radial) and $\theta$ (tangential) to characterize the stress $\sigma$. The axisymmetric equation of equilibrium then reads \cite{auckland}:

    \begin{eqnarray}
    \frac{\partial \sigma_{rr}}{\partial r} + \frac{1}{r} \frac{\partial \sigma_{r\theta}}{\partial \theta} + \frac{1}{r}(\sigma_{rr} - \sigma_{\theta \theta}) = 0 \nonumber \\
    \frac{\partial \sigma_{r\theta}}{\partial r} + \frac{1}{r} \frac{\partial \sigma_{\theta \theta}}{\partial \theta} + \frac{2\sigma_{r\theta}}{r} = 0
    \label{eq1}
    \end{eqnarray}
    
    \noindent
    where $\sigma_{ij}$ are the different components of the stress tensor. If the flywheel of mass $m$, is rotating at constant angular velocity $\omega$, a point at distance $r$ from the rotation axis, experiences the radial centripetal force $F = - mr\omega^2$, and no tangential force, so Eq.~\ref{eq1} becomes:
    
    \begin{eqnarray}
    \frac{\partial \sigma_{rr}}{\partial r}  + \frac{1}{r}(\sigma_{rr} - \sigma_{\theta \theta}) = - \rho r \omega^2
    \label{eq2}
    \end{eqnarray}
    
    \noindent Then, using Hooke's law for isotropic materials deformation under stress, the relationship between components of the stress $\sigma$ and the strain $\epsilon$ may be written as:
    
    \begin{eqnarray}
    \left(\begin{array}{cc}
    \sigma_{rr} & \sigma_{r\theta}\\ 
    \sigma_{r\theta} & \sigma_{\theta\theta}
    \end{array} \right) = \frac{E}{1-\nu^2}\left[(1-\nu) \left( \begin{array}{cc}
    \epsilon_{rr} & \epsilon_{r\theta}\\ 
    \epsilon_{r\theta} & \epsilon_{\theta\theta}
    \end{array} \right) + \nu \mathds{1}(\epsilon_{rr} + \epsilon_{\theta\theta})\right]
    \label{eq3}
    \end{eqnarray}
    
    \noindent where $E$ is Young's modulus and $\nu$ is Poisson's ratio. Further, the strain-displacement equations read: $\epsilon_{rr} = \frac{\partial u_r}{\partial r}$ and $\epsilon_{\theta\theta} = \frac{1}{r}\frac{\partial u_{\theta}}{\partial \theta} + \frac{u_r}{r} = \frac{u_r}{r}$; so, combination of these and Eqs.~(\ref{eq2}) and (\ref{eq3}), yields the equation for the displacement:
    
    \begin{eqnarray}
    \frac{d^2u_r}{dr^2} + \frac{1}{r} \frac{du_r}{dr} - \frac{1}{r^2} u_r = - \frac{1-\nu^2}{E}\rho r \omega^2
    \label{eq4}
    \end{eqnarray}
    
    \noindent which is an ordinary differential equation (ODE) that we can solve using the ansatz $u_r = \alpha r^{\beta}$:
    
    \begin{eqnarray}
    u_r = k_1 r + k_2 r^{-1} - \frac{1-\nu^2}{8E}\rho \omega^2 r^3
    \label{eq5}
    \end{eqnarray}
    
    \noindent where the constants $k_1$ and $k_2$ are constrained by the boundary conditions, given by the geometry of the flywheel. Then, using the strain-displacement equations and Eq. (\ref{eq3}), we get the stress equations:
    
    \begin{eqnarray}
    \sigma_{rr} = \Tilde{k}_1 - \Tilde{k}_2 \frac{1}{r^2} - (3+\nu)\frac{\rho \omega^2}{8}r^2 \nonumber \\
    \sigma_{\theta \theta} = \Tilde{k}_1 + \Tilde{k}_2 \frac{1}{r^2} - (1+3\nu)\frac{\rho \omega^2}{8}r^2
    \label{eq6}
    \end{eqnarray}
    
    \noindent where $\Tilde{k}_1 = \frac{E}{1-\nu}k_1$ and $\Tilde{k}_2 = \frac{E}{1+\nu}k_2$.\\
    
    Now that we derived the basic formulae, we can now analyze them considering two important geometrical configurations: solid disk and ring, and see what information we may obtain as we consider acceleration and deceleration of the flywheel. 
    
    \subsubsection{Solid disk}
\begin{figure}
    \centering
    \resizebox{0.75\textwidth}{!}{\includegraphics{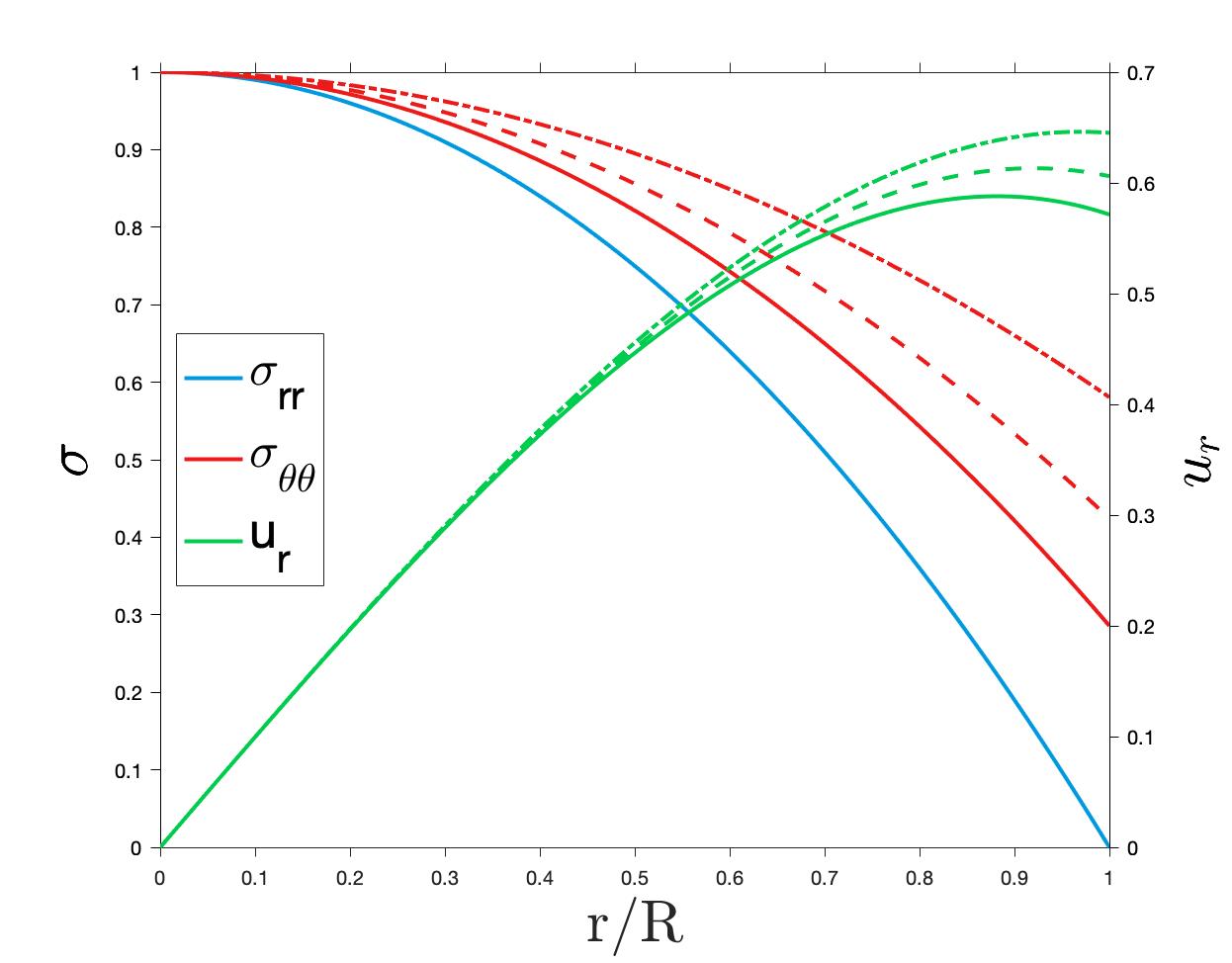}}
    \caption{Stress matrix elements $\sigma_{rr}$ and $\sigma_{\theta\theta}$ (left axis), and displacement $u_r$ (right axis) for different Poisson ratios: $\nu = 0.1$ (dashed-dotted lines), $\nu = 0.3$ (dashed lines), $\nu = 0.5$ (solid lines). Stress and displacement are normalized to $\frac{8}{(3+\nu) \rho \omega^2 R^2}$ and $\frac{8E}{(3+\nu)(1-\nu) \rho \omega^2 R^3}$ respectively. The scaled stress component $\sigma_{rr}$ does not change with $\nu$.} 
    \label{fig:harddisk}
    \end{figure}

    For a solid disk of radius $R$, $\Tilde{k}_2$ in Eq.~(\ref{eq6}) has to be set to zero to avoid the rapid divergence of the $\Tilde{k}_2 r^{-2}$, for $r\rightarrow 0$; and since the normal stress is always zero at a free boundary, the radial stress goes to zero at the edge of the disk, $\sigma_{rr} (R) = 0$. We thus find that $\Tilde{k}_1 = (3+\nu)\frac{\rho \omega^2}{8}R^2$, and we end up with the following expressions of the stress and displacement for the solid disk geometry:
    
    \begin{eqnarray}
    \sigma_{rr} = (3+\nu)\frac{\rho \omega^2}{8}(R^2 - r^2) \nonumber \\
    \sigma_{\theta \theta} = (3+\nu)\frac{\rho \omega^2}{8}(R^2 - \frac{1+3\nu}{3+\nu}r^2) \nonumber \\
    u_r = (3+\nu)\frac{\rho \omega^2}{8}\frac{1-\nu}{E}(R^2 - \frac{1+\nu}{3+\nu}r^2)r
    \label{eq7}
    \end{eqnarray}
    
    \noindent which are shown in Fig. \ref{fig:harddisk} as functions of the ratio $r/R$. The displacement is strictly zero at the center of the disk, and, as it is non-uniform, naturally reaches a maximum before and not at the edge of the disk (otherwise this could lead to dislocation). Stress is maximum at $r=0$ and, as expected, decreases with increasing $r$. At its maximum, the stress is $\sigma_{rr} = \sigma_{\theta\theta} = (3+\nu)\frac{\rho \omega^2 R^2}{8}$. As the Poisson decreases, both the maximum displacement ($u_r$) and the minimum hoop stress ($\sigma_{\theta\theta}$) increase. The scaled radial stress ($\sigma_{rr}$) does not change with $\nu$ as it only appears in the overall prefactor of $u(r)$. 
    
    \subsubsection{Hollow disk (ring)}
    For a hollow disk of inner radius $R_1$ and outer radius $R_2$, we have two free boundaries and hence two boundary conditions: the radial component of the stress must be zero at both $r=R_1$ and $r=R_2$. We thus have to solve the system of equations for $\Tilde{k}_1$ and $\Tilde{k}_2$:
    
    \begin{eqnarray}
    \Tilde{k}_1 - \Tilde{k}_2 \frac{1}{R_1^2} - (3+\nu)\frac{\rho \omega^2}{8}R_1^2 = 0 \nonumber \\
    \Tilde{k}_1 - \Tilde{k}_2 \frac{1}{R_2^2} - (1+3\nu)\frac{\rho \omega^2}{8}R_2^2 = 0
    \label{eq8}
    \end{eqnarray}
    
    \noindent We then find that $\Tilde{k}_1 = - \frac{3+\nu}{8}\rho \omega^2 R_1^2 R_2^2$ and $\Tilde{k}_2 = \frac{3+\nu}{8}\rho \omega^2 (R_1^2 + R_2^2)$, and the radial stress, hoop stress, and displacement read:
    
    \begin{eqnarray}
    \sigma_{rr} = (3+\nu)\frac{\rho \omega^2}{8}(R_1^2 + R_2^2 - r^2 - \frac{R_1^2 R_2^2}{r^2}) \nonumber \\
    \sigma_{\theta \theta} = (3+\nu)\frac{\rho \omega^2}{8}(R_1^2 + R_2^2  - \frac{1+3\nu}{3+\nu}r^2 + \frac{R_1^2 R_2^2}{r^2}) \nonumber \\
    u_r = (3+\nu)\frac{\rho \omega^2}{8}\frac{1-\nu}{E}(R_1^2 + R_2^2  - \frac{1+\nu}{3+\nu}r^2 + \frac{1+\nu}{1-\nu}\frac{R_1^2 R_2^2}{r^2})r
    \label{eq9}
    \end{eqnarray}
    
\begin{figure}
    \centering
    \resizebox{0.75\textwidth}{!}{\includegraphics{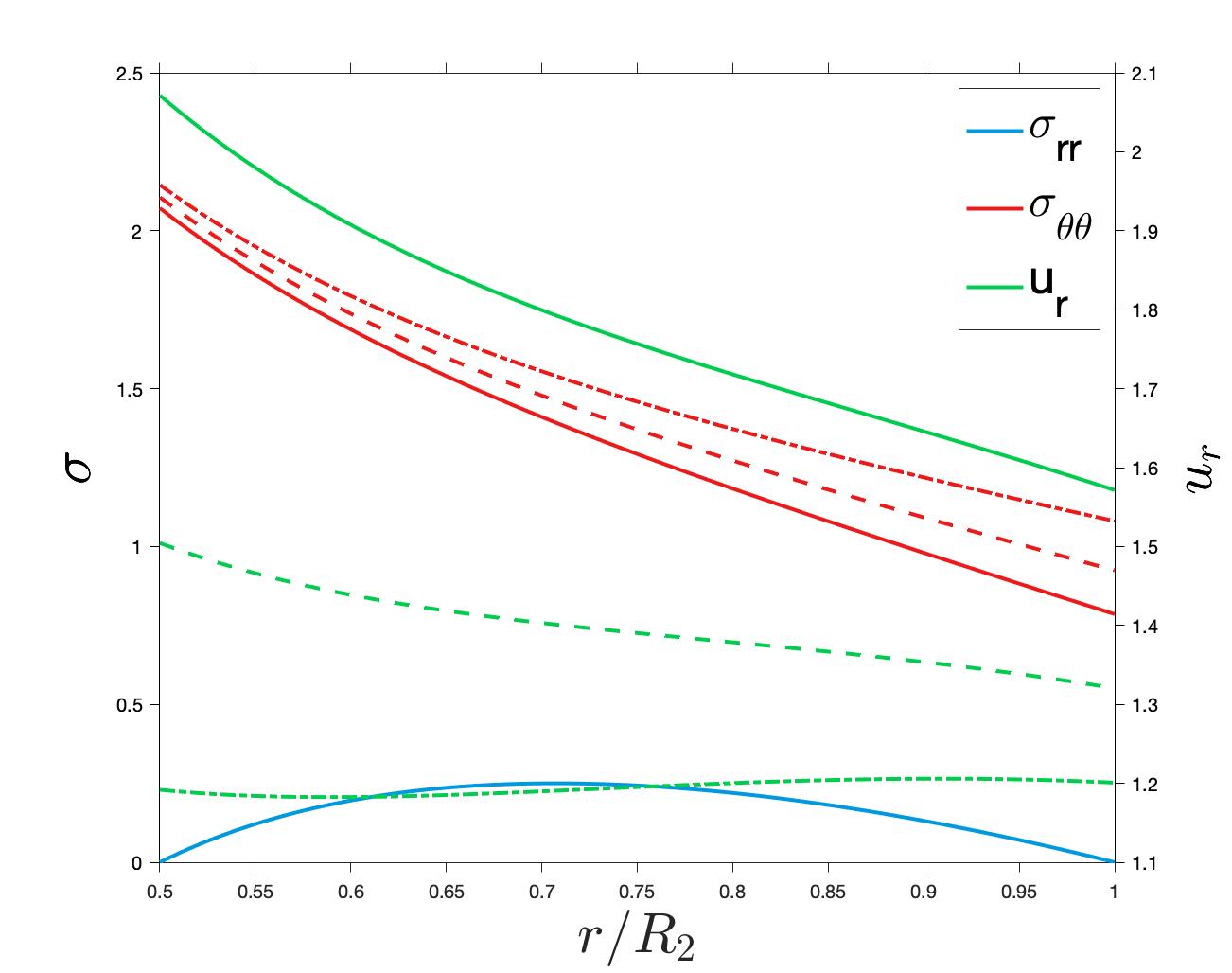}}
    \caption{Stress matrix elements $\sigma_{rr}$ and $\sigma_{\theta\theta}$ (left axis), and displacement $u_r$ (right axis) for different Poisson ratios: $\nu = 0.1$ (dashed-dotted lines), $\nu = 0.3$ (dashed lines), $\nu = 0.5$ (solid lines) and for the inner-outer radius ratio of $R_1 / R_2 = 0.5$. Stress and displacement are normalized to $\frac{8}{(3+\nu) \rho \omega^2 R^2}$ and $\frac{8E}{(3+\nu)(1-\nu) \rho \omega^2 R_2 ^3}$ respectively.}
    \label{fig:hollowdisk}
    \end{figure}

    \noindent Equations (\ref{eq9}) reduce to Eqs. (\ref{eq7}) if we let $R_1$ go to $0$. The stress components $\sigma_{rr}$ and $\sigma_{\theta\theta}$, and the displacement $u_r$ are shown in Fig. \ref{fig:hollowdisk}. The component $\sigma_{rr}$ is zero at both $r=R_1$ and $r=R_2$, but $\sigma_{\theta \theta}$ is maximum at $r=R_1$ with a value of $\sigma_{\theta\theta} = (3+\nu)\frac{\rho \omega^2}{8}(R_2^2  + \frac{1-\nu}{3+\nu}R_1^2)$, and decreases as $r$ increases. As $\nu$ increases, hoop stress decreases, but displacement sharply increases. 
    
    \subsubsection{Shape factor}
    We may now calculate the shape factor shown in Eq. (\ref{eq:spece}). Starting from:
    
    \begin{eqnarray}
    e = W_{\mathrm{max}}/M = \frac{J\omega_{\mathrm{max}}^2}{2M}
    \label{eq:shape}
    \end{eqnarray}
    
    \noindent where $W_{\mathrm{max}}$ is the maximum stored kinetic energy, $J$ the moment of inertia and $M$ the mass of the flywheel, we make explicit the maximum stress for a hollow disk (or cylinder) and the moment of inertia of a hollow cylinder, to obtain the specific energy as follows:
    
    \begin{eqnarray}
    e = \frac{\frac{1}{2}M(R_1^2 + R_2^2)\frac{8\sigma_u}{\rho(3+\nu)(R_2^2  + \frac{1-\nu}{3+\nu}R_1^2)}}{2M} = \frac{2(R_1^2 + R_2^2)}{(3+\nu)(R_2^2  + \frac{1-\nu}{3+\nu}R_1^2)} \frac{\sigma_u}{\rho}
    \label{eq:shape2}
    \end{eqnarray}
    
    \noindent By comparison with Eq.~(\ref{eq:spece}), we find that $K$ reads
    
    \begin{eqnarray}
    K = \frac{2(R_1^2 + R_2^2)}{(3+\nu)(R_2^2  + \frac{1-\nu}{3+\nu}R_1^2)}
    \label{eq:shape3}
    \end{eqnarray}
    
    \noindent for hollow and solid disks (for which the inner radius is simply set to zero $R_1 = 0$), and cylinders for which we simply increase the length along the $z$-axis, leaving the properties in the $\hat{\theta}$ and $\hat{r}$ directions unchanged. The theoretical maximum value of $K$ is $1$ for $R_1=R_2$, for a perfectly flat ring.

\subsection{Acceleration and deceleration}
    As energy is stored in the flywheel or extracted from it, an extra force is added on the flywheel to either accelerate or decelerate its rotation. This force being along the tangential ($\hat{\theta}$) direction and with magnitude $F = ma = mr\frac{d\omega}{dt}$, Eq. (\ref{eq1}) may be rewritten as follows:

    \begin{eqnarray}
    \frac{\partial \sigma_{rr}}{\partial r} + \frac{1}{r} \frac{\partial \sigma_{r\theta}}{\partial \theta} + \frac{1}{r}(\sigma_{rr} - \sigma_{\theta \theta}) = - \rho r \omega^2 \nonumber\\
    \frac{\partial \sigma_{\theta r}}{\partial r} + \frac{1}{r} \frac{\partial \sigma_{\theta \theta}}{\partial \theta} + \frac{2\sigma_{r\theta}}{r} = \rho r \frac{d\omega}{dt}
    \label{eq10}
    \end{eqnarray}
    
    \noindent where $\frac{d\omega}{dt}$ is positive when energy is added to the flywheel, and negative when it is extracted. Equation (\ref{eq2}) is essentially a special case of Eqs. (\ref{eq10}) with $\frac{d\omega}{dt} = 0$. Since the kinetic energy of the flywheel is $E = \frac{1}{2}J\omega^2$ and the power is $P=\frac{dE}{dt}$, we have $\frac{d\omega}{dt} = \frac{P}{J\omega}$. We then use Eq. (\ref{eq3}) and the strain-displacement relations:
    
    \begin{eqnarray}
    \epsilon_{rr} = \frac{\partial u_r}{\partial r} \nonumber \\
    \epsilon_{\theta \theta} = \frac{1}{r}\frac{\partial u_{\theta}}{\partial \theta} +  \frac{u_r}{r} \nonumber \\
    \epsilon_{r \theta} = \frac{\partial u_{\theta}}{\partial r} + \frac{1}{r}\frac{\partial u_r}{\partial \theta} - \frac{u_{\theta}}{r}
    \label{eq11}
    \end{eqnarray}
    
    \noindent to get the system of partial differential equations that we need to solve for:
    
    \begin{eqnarray}
    r^2 \frac{\partial^2 u_r}{\partial r^2} + (1-\nu)\frac{\partial^2 u_r}{\partial \theta^2} + r \frac{\partial u_r}{\partial r} - u_r 
    + r\frac{\partial^2 u_{\theta}}{\partial r \partial \theta} - (2-\nu) \frac{\partial u_{\theta}}{\partial \theta} = - \frac{(1-\nu^2)\rho\omega^2}{E} r^3 \nonumber \\
    r^2 \frac{\partial^2 u_{\theta}}{\partial r^2} + \frac{1}{1-\nu}\frac{\partial^2 u_{\theta}}{\partial \theta^2} + r \frac{\partial u_{\theta}}{\partial r} - u_{\theta} 
    + r\frac{1}{1-\nu}\frac{\partial^2 u_r}{\partial r \partial \theta} + \frac{2-\nu}{1-\nu}\frac{\partial u_r}{\partial \theta} = \frac{(1+\nu)\rho P}{J\omega E}r^3
    \label{eq12}
    \end{eqnarray}
    
    Since we assumed from the outset symmetry along the $z$-axis, $u_r$ and $u_{\theta}$ do not depend on $\theta$ and we can set to zero all the terms with derivatives with respect to $\theta$, which yields:
    
    \begin{eqnarray}
    r^2 \frac{d^2 u_r}{d r^2} + r \frac{d u_r}{d r} - u_r = - \frac{(1-\nu^2)\rho\omega^2}{E} r^3 \nonumber \\
    r^2 \frac{d^2 u_{\theta}}{d r^2} + r \frac{d u_{\theta}}{d r} - u_{\theta} = \frac{(1+\nu)\rho P}{J\omega E}r^3
    \label{eq13}
    \end{eqnarray}
    
    \noindent whose solutions can be written as follows:
    
    \begin{eqnarray}
    u_r = \alpha r^3 + U_r = k_1 r + k_2 r^{-1} - \frac{1-\nu^2}{8E}\rho \omega^2 r^3 \nonumber \\
    u_{\theta} = \beta r^3 + U_{\theta} = k_3 r + k_4 r^{-1} + \frac{1+\nu}{8E} \frac{\rho P}{J\omega} r^3
    \label{eq14}
    \end{eqnarray}
    
    \noindent  and using Eqs. (\ref{eq3}) and (\ref{eq11}), we find that the stress matrix elements read:
    
    \begin{eqnarray}
    \sigma_{rr} = \Tilde{k}_1 - \Tilde{k}_2 \frac{1}{r^2} - (3+\nu)\frac{\rho \omega^2}{8}r^2 \nonumber \\
    \sigma_{\theta \theta} = \Tilde{k}_1 + \Tilde{k}_2 \frac{1}{r^2} - (1+3\nu)\frac{\rho \omega^2}{8}r^2 \nonumber \\
    \sigma_{\theta r} = - \Tilde{k}_4 \frac{1}{r^2} + \frac{\rho P}{4J\omega} r^2
    \label{eq15}
    \end{eqnarray}
    
    \noindent where $\Tilde{k}_1 = \frac{E}{1-\nu}k_1$, $\Tilde{k}_2 = \frac{E}{1+\nu}k_2$, and $\Tilde{k}_4 = \frac{2E}{1+\nu}k_4$. Note the radial displacement $u_r$ is the same as in Eq. (\ref{eq5}) for the constant angular velocity rotation.
    
    \subsubsection{Solid disk}
\begin{figure}
    \centering
    \resizebox{0.75\textwidth}{!}{\includegraphics{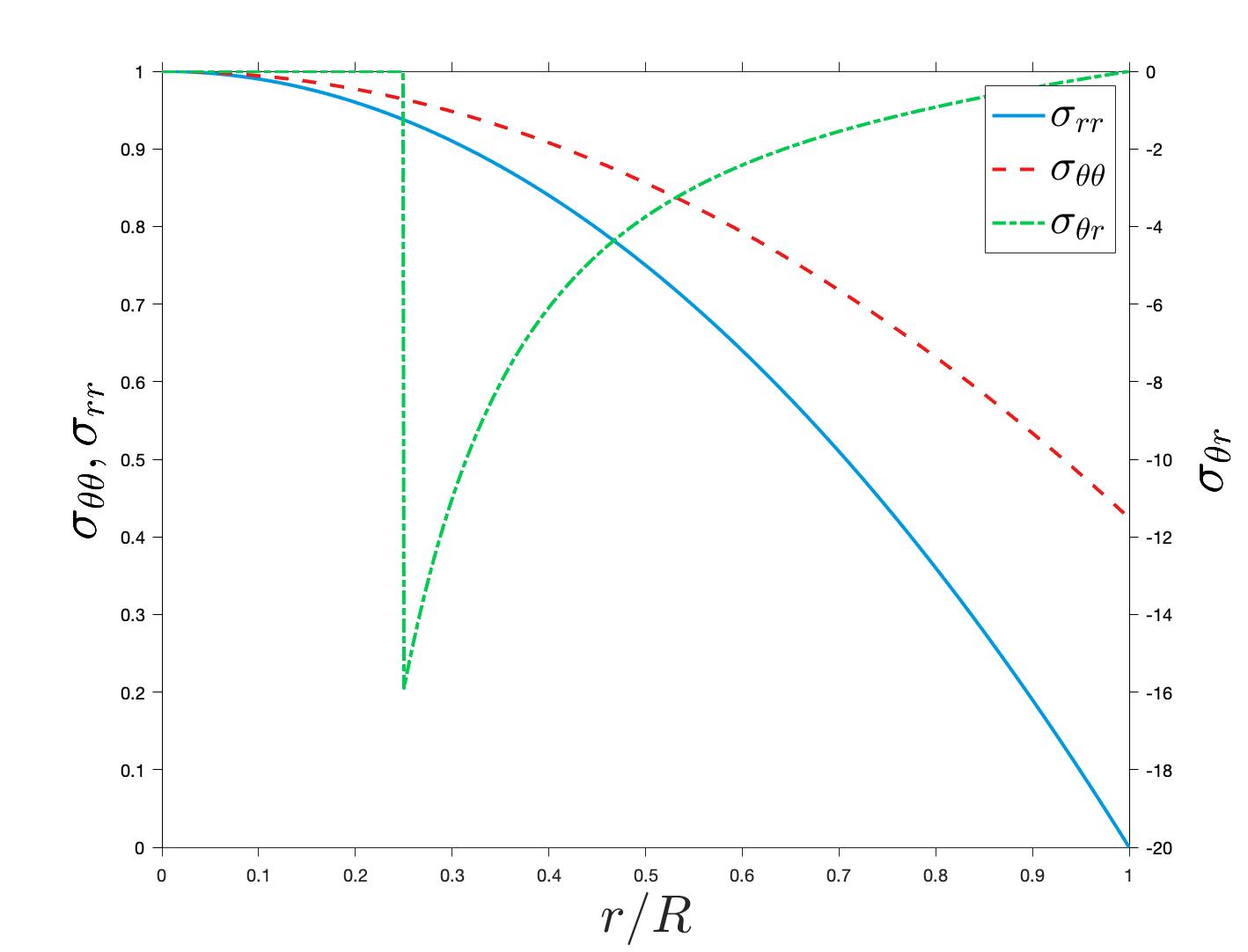}}
    \caption{Plots of $\sigma_{rr}$, $\sigma_{\theta\theta}$ (left axis) and $\sigma_{\theta r}$ (right axis) for Poisson's ratio of $\nu = 0.3$. The stresses $\sigma_{rr}$ and $\sigma_{\theta\theta}$ are normalized to $\frac{8}{(3+\nu) \rho \omega^2 R^2}$ and $\sigma_{\theta r}$ is normalized to $\frac{\rho P}{4J\omega} R^2$. The shaft radius is $r_{\rm shaft} = 0.25 R$.}
    \label{fig:accdisk}
    \end{figure}

    Using the same boundary conditions as for the constant-angular-velocity rotation case above, we find: $\Tilde{k}_2 = 0$ and $\Tilde{k}_1 = (3+\nu)\frac{\rho \omega^2}{8}R^2$. However, the shear stress component $\sigma_{\theta r}$ requires some attention. The shear stress being zero at a free boundary, the boundary condition simply reads $\sigma_{\theta r} = 0$ at $r = R$, which gives $\Tilde{k}_4 = \frac{\rho P}{4J\omega} R^4$. This seems problematic at first, since it would mean that the shear stress goes to infinity at $r=0$. However, we have to keep in mind that in order to accelerate or decelerate the cylinder, a torque has to be applied somewhere. In our case this torque is applied on a shaft of smaller radius ($r_{\mathrm{shaft}}$) than the flywheel itself. Therefore, within this radius, the shear stress is zero because of the uniformly applied torque; beyond this radius, the shear stress follows Eq. (\ref{eq15}). If we let the radius of the shaft go to zero, applying a nonzero torque would require an infinite force, hence the diverging behaviour of the shear stress at $r=0$ if $r_{\mathrm{shaft}}=0$. The stress tensor elements may thus read: 
    
    \begin{eqnarray}
    \sigma_{rr} = (3+\nu)\frac{\rho \omega^2}{8}(R^2 - r^2) \nonumber \\
    \sigma_{\theta \theta} = (3+\nu)\frac{\rho \omega^2}{8}(R^2 - \frac{1+3\nu}{3+\nu}r^2) \nonumber \\
    \sigma_{\theta r} =\begin{cases}
    0, & \text{if $r<r_{\mathrm{shaft}}$} \\
    \frac{\rho P}{4J\omega}(r^2 -\frac{R^4}{r^2}), & \text{if $r>r_{\mathrm{shaft}}$}
    \end{cases}
    \label{eq:finally}
    \end{eqnarray}
    
    \noindent and they are shown in Fig. \ref{fig:accdisk}. The radial and hoop stresses  $\sigma_{rr}$ and  $\sigma_{\theta \theta}$ are exactly the same as for the constant angular velocity case. The maximum value of the shear stress $\sigma_{\theta r}$ is $\frac{\rho P}{4J\omega}\left(r_{\mathrm{shaft}}^2 -\frac{R^4}{r_{\mathrm{shaft}}^2}\right)$ at $r=r_{\mathrm{shaft}}$.
    
    \subsubsection{Hollow disk}
    In the accelerating hollow disk, the radial and hoop stresses $\sigma_{rr}$ and $\sigma_{\theta \theta}$, as with the solid disk, do not change from the constant angular velocity case and take the forms of Eqs.~(\ref{eq9}). The shear stress $\sigma_{\theta r}$, however, takes the form of Eq.~(\ref{eq:finally}) with the appropriate shaft radius.
    
    \subsubsection{Comparing radial and hoop stress with shear stress}
    The mathematical expressions of the stress components in the case of rotation with acceleration may be simplified using the formulae obtained for the constant-angular-velocity rotation case if $\sigma_{\theta r} \ll \sigma_{rr},\sigma_{\theta \theta}$. Leaving the radial dependencies out, we can see that $\sigma_{\theta r} \sim \frac{\rho P}{J\omega}$ and that both $\sigma_{rr}$ and $\sigma_{\theta \theta}$ behave as $\rho \omega^2$, so that $\frac{\rho P}{J\omega} \ll \rho \omega^2$ or, equivalently, $\frac{P}{\omega} \ll J\omega^2 = W$, where $W$ is the rotational kinetic energy of the flywheel. Hence, for sufficiently high kinetic energy and angular velocity, the shear stress $\sigma_{\theta r}$ may be neglected. However, since $\sigma_{\theta r} \sim \frac{1}{\omega}$, at small angular velocities, the shear stress can reach quite high values. In practice, this is avoided by keeping the flywheel rotating above a certain minimum value $\omega_{\mathrm{min}}$, typically around $1/2$ and $1/3$ of $\omega_{\mathrm{max}}$, the maximum angular velocity, so that the shear stress does not become critically large for a given power \cite{joule}.
%%%%%%%%%%%%%%%%%%%%%%%%%%%%%%%%%%%%%%%%%%%%%%%%%%%%%%%%%%%%%%%%%%%%%%%%%%%%%%%%%%%%%%%%%%%%%%%%%%%%%%%%%%%%
%%%%%%%%%%%%%%%%%%%%%%%%%%%%%%%%%%%%%%%%%%%%%%%%%%%%%%%%%%%%%%%%%%%%%%%%%%%%%%%%%%%%%%%%%%%%%%%%%%%%%%%%%%%%
%%%%%%%%%%%%%%%%%%%%%%%%%%%%%%%%%%%%%%%%%%%%%%%%%%%%%%%%%%%%%%%%%%%%%%%%%%%%%%%%%%%%%%%%%%%%%%%%%%%%%%%%%%%%
%%%%%%%%%%%%%%%%%%%%%%%%%%%%%%%%%%%%%%%%%%%%%%%%%%%%%%%%%%%%%%%%%%%%%%%%%%%%%%%%%%%%%%%%%%%%%%%%%%%%%%%%%%%%
%%%%%%%%%%%%%%%%%%%%%%%%%%%%%%%%%%%%%%%%%%%%%%%%%%%%%%%%%%%%%%%%%%%%%%%%%%%%%%%%%%%%%%%%%%%%%%%%%%%%%%%%%%%%
\section{Electrical design of FES systems}
\begin{figure}
    \centering
    \resizebox{0.75\textwidth}{!}{\includegraphics{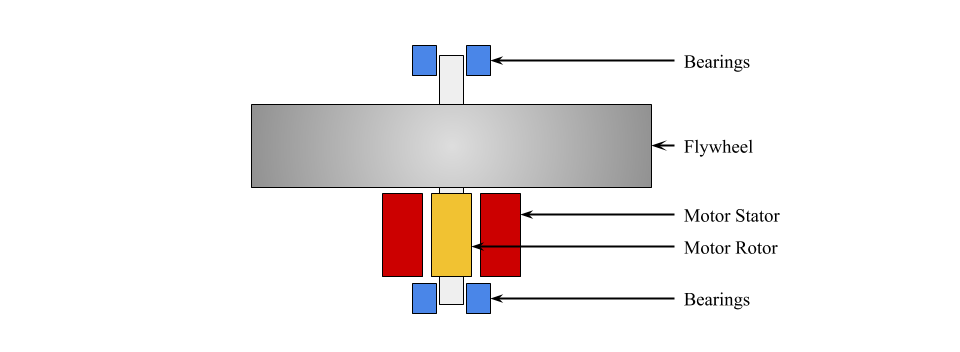}}
    \caption{Schematic of the electro-mechanical design of a FES system.}
    \label{fig:flywheel}
    \end{figure}

\begin{figure}
    \centering
    \resizebox{0.75\textwidth}{!}{\includegraphics{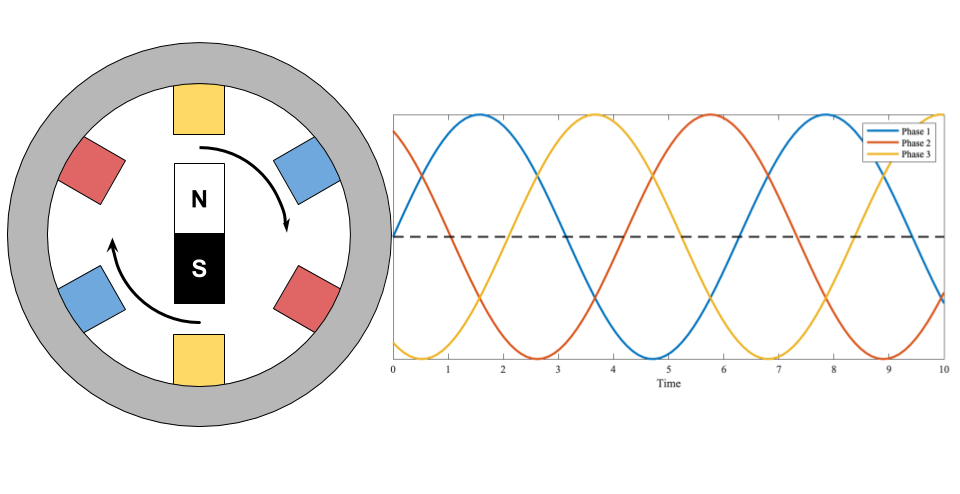}}
    \caption{Schematic of a 3-phase AC induction motor and the corresponding voltage waveforms of the different coils.}
    \label{fig:3phase}
    \end{figure}

    The electrical part of a flywheel energy storage system, as shown in Fig.~\ref{fig:flywheel}, has an electrical machine connected to it which can operate on both motoring and generating modes; the motoring mode during charging of the flywheel, and the generating mode during discharging of the flywheel. The flywheel is mechanically coupled with the rotor of the electrical machine and supports are provided by two bearings at the top and the bottom of the arrangement. The rotor of the machine holds a steady magnetic field, the rotation of which generates a three-phase alternating voltage at the terminals of the stator during the discharging phase (see Fig.~\ref{fig:3phase}). During the charging mode, an electrical current fed to the machine terminals creates a magnetic field in the stator winding, which interlocks with the rotor magnetic field causing the rotation of the rotor-flywheel couple and hence charges the flywheel. Also, the interlocking of rotor and flywheel leads to the whole system's moment of inertia to be defined as the sum of the moments of inertia of the flywheel and rotor respectively: $J = J_{\rm f} + J_{\rm r}$.
    
    % use hat for per-unit

    \subsection{Charging, storage, and discharge phases of flywheels}
\begin{figure}
        \centering
        \resizebox{0.75\textwidth}{!}{\includegraphics{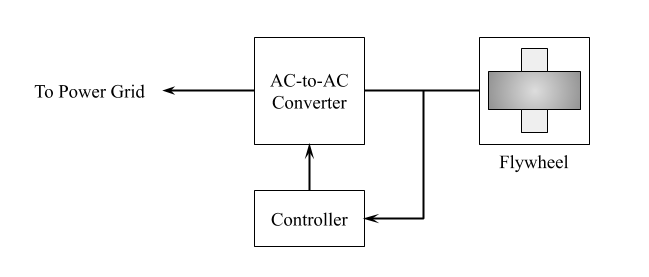}}
        \caption{Basic system circuit diagram of a FESS. The line to the power grid has constant voltage and frequency, whereas the line between the AC-to-AC converter and the flywheel has variable voltage and frequency. The controller takes real-time information from the signal going to the flywheel and adjusts the AC-to-AC converter accordingly.}
        \label{fig:fwcircdgm}
    \end{figure}

    A flywheel typically has three distinct operation phases. First, when the electrical machine is in motoring mode, the flywheel charges and its speed goes from some minimal speed (which can be 0) up to its rated speed $\omega_{\rm c}(t_{\rm c})$ rad$\cdot$s$^{-1}$ over its charging time $t_{\rm c}$, after which the electrical supply is cut off. In an induction machine driven flywheel system, the speed is easily controlled via the supplied voltage \cite{chapman2012synchronous}; in the case of a synchronous machine, as considered in the present analysis, the speed is always constant for a given frequency \cite{chapman2012synchronous}. Note that it is neither possible, nor practical to plug in an uncharged flywheel directly to the grid because it cannot pick-up its speed instantaneously and this would anyway affect in a detrimental fashion the dynamics of the electrical grid. Hence, during the charging phase, a controlled input from the AC-AC converter (see Fig. \ref{fig:fwcircdgm}) is provided to the electrical machine to ensure a gradual rise of the flywheel rotation speed. After it has been charged, the flywheel is in the storage phase, where the electrical machine is not operating, but the speed of the flywheel decays slowly due to the friction with the bearings and with ambient air. The speed before the discharging phase is $\omega_{\rm s}(t_{\rm s})$ (which is smaller than $\omega_{\rm c}$) where $t_{\rm s}$ is the storage time. During the third phase, i.e. the discharging phase of the flywheel, the initial speed of the flywheel is $\omega_{\rm s}(t_{\rm s})$, from which the speed slowly decays. But in contrast to the storage phase where the speed decay is only due to friction, the speed decay is affected by the electrical speed $\omega_{\rm e}(t)$ (which arises from the magnetic interaction of the rotor and the stator of the electrical machine) in addition to the frictional decay. Hence, at any instant $t$ of the discharging phase, the rotation speed of flywheel can be written as:
    \begin{equation}
    %\begin{split}
        %\omega_{d}(t) &= \omega_{s}(t_{s}) - B\omega_{d}(t-\Delta t) - \omega_{e}(t)\\
        %\text{or, }\omega_{d}(t) &= \omega_{s}(t_{s}) - B\{\omega_d(t)+\alpha(t)\Delta t\} - \omega_{e}(t)\\
        %\text{or, }
        \omega_{\rm d}(t) = \frac{\omega_{\rm s}(t_{\rm s}) - B\alpha(t)\Delta t - \omega_{\rm e}(t)}{(1+B)}\\
    %\end{split}
    \label{eq:DischargeOmegaforSpeed}
    \end{equation}
    where $B$ is the equivalent friction coefficient of the two bearings, $\alpha(t)$ is the instantaneous acceleration, and $\Delta t$ is a short time interval for acceleration measurement or approximation: $\alpha(t) = \frac{\omega_{\rm d}(t)-\omega_{\rm d}(t+\Delta t)}{\Delta t}$. As the machine operates at its synchronous speed, which itself decays over time, the level of electrical current generated consequently decays over time. In order to maintain the synchronization with the grid, the AC-AC converter (Fig.~\ref{fig:fwcircdgm}) is used to convert the decaying-frequency output of the flywheel to constant frequency constant voltage electrical power output.

    %\textcolor{red}{Based on the equations above, and the dynamics of synchronous machine at different speed input or different frequency input, the following speed-time curve is constructed. For simplicity, the frequency at charging phase is taken to be raised linearly. But this is entirely controllable by the user from the AC-AC controller in a real-life FESS.}
    
    \subsection{Torque balance in the flywheel}
\begin{figure}[h]
    \centering
    \includegraphics[width = 0.8\textwidth]{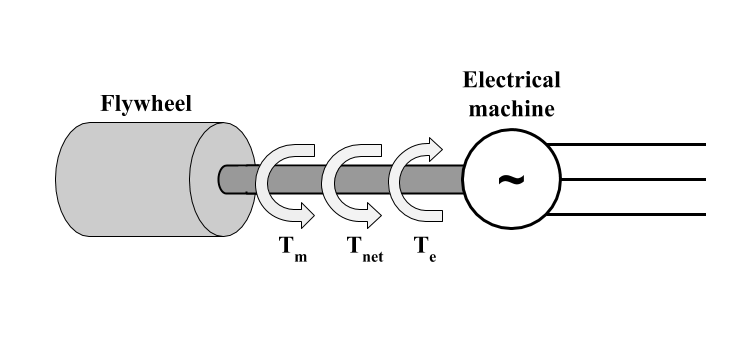}
    \caption{Flywheel electrical diagram: torque balance in the flywheel energy storage system.}
    \label{fig:FlywheelElectrical}
    \end{figure}

    While the flywheel is being discharged, the electrical machine is in generating mode, and an electrical current is drawn from it. According to Lenz's law, this current opposes its cause and creates an opposing torque known as electrical torque \cite{chapman2012synchronous}. Because of this opposing torque, as shown in Fig. \ref{fig:FlywheelElectrical}, the flywheel actually operates on a net torque which is the difference between the mechanical and the electrical torques: $T_{\rm net} = T_{\rm m}-T_{\rm e}$.
    
    The net torque is related to the moment of inertia $J$, and reads:
    \begin{equation}
    T_{\rm net} = J\frac{{\rm d}\omega_{\rm d}(t)}{{\rm d}t} = \frac{2H}{\omega_{\rm d_r}^2S_{\rm r}}\frac{{\rm d}\omega_{\rm d}(t)}{{\rm d}t}
    \label{eq:nettorque}
    \end{equation}
    where $H$ is the system's inertia constant defined as the ratio of the rated kinetic energy of the flywheel-rotor couple to the electrical power output of the machine: $H = KE_{\rm r}/S_{\rm r} = J\omega_{{\rm d_r}}^2/2S_{\rm r}$, with $\omega_{{\rm d_{r}}}$ being the flywheel rated speed during the discharge phase in rad$\cdot$s$^{-1}$, and $S_{\rm r}$ being the rated output of the electrical machine in VA. Note that the constant $2H$, often written as $M$, is the mechanical starting time of the entire flywheel-electrical generator system \cite{kundur1994power}, i.e. the time taken by an electrical machine to get to its rated speed from zero up to the rated level of mechanical power input. %Hence, using this notion, it is easy to measure $M=2H$ and plug in the constant value rather than proceeding analytically.

    % We know, power output = torque $\times$ angular speed. i.e. $S_{rated} = T_{rated}\times\omega_{rated}$. Hence, $T_{rated} = S_{rated}/\omega_{d_r}$. So, we can rewrite the torque equation as,
    %clarify omega_d_r
    % \begin{equation}
    %     T_{net} = \frac{2H}{\omega_{d_{r}}^2}S_{rated}\frac{d\omega_{d}(t)}{dt}
    %     \label{eq:nettorque2}
    % \end{equation}
    
    \noindent Rearranging the above expression and writing the time-varying terms in per-unit notation (here denoted with an overhead bar: $\overline{\phantom{\omega}}$), we obtain the first-order differential equation:
    %\begin{equation}
    %\begin{split}
    %    \frac{T_{net}}{S_{rated}/\omega_{d_r}} &= %2H\frac{d}{dt}\left(\frac{\omega_d(t)}{\omega_{d_r}}\right)\\
    %    \text{or, }\overline{T}_{net} &= %2H\frac{d\overline{\omega}_d(t)}{dt}
    %    \label{eq:nettorquepu}
    %\end{split}
    %\end{equation}
    %We can write this differential equation as,
    \begin{equation}
        \Dot{\overline{\omega}}_{\rm d}(t) = \frac{\overline{T}_{\rm net}}{2H} = \frac{\overline{T}_{\rm m}-\overline{T}_{\rm e}}{2H}
        \label{eq:OmegaDiffEq}
    \end{equation}
    satisfied by the flywheel rated rotation speed during the discharge phase. 
    % electrical torque for charging phase as well (from electrical)
    
    \subsection{Electrical torque equation}
\begin{figure}[h]
        \centering
        \includegraphics[width = .5\textwidth]{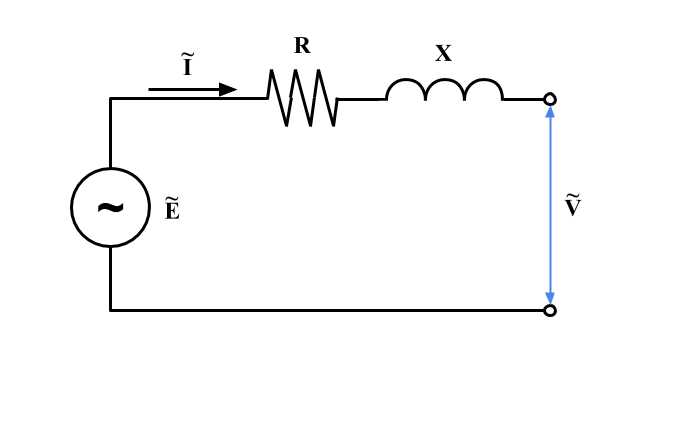}
        \caption{Electrical machine internal diagram. It could assumed as a single-phase machine or single-phase equivalent of a three-phase machine. For a three-phase system, the power output is three times that of a single-phase system.}
        \label{fig:electricalSLD}
    \end{figure}

    As shown in Fig. \ref{fig:electricalSLD}, an electrical generator can simply be modeled as an ideal voltage source $\Tilde{E}$ connected to its internal impedance given as $Z=R+jX = |Z|\angle\theta$, with $R$ being the electrical resistance, $X$ being the reactance, and with $j^2=-1$. Denoting the terminal voltage $\Tilde{V}$, at the terminals of the machine, and the electrical current that flows through the machine, $\Tilde{I}$, then the apparent power delivered is given by:
    \begin{equation}
        \Tilde{S} = \Tilde{V}\Tilde{I}^*
        \label{eq:apparentpower}
    \end{equation}
    
\begin{figure}[h]
        \centering
        \includegraphics[trim={10cm 14cm 8cm 2.5cm},clip,width = .5\textwidth]{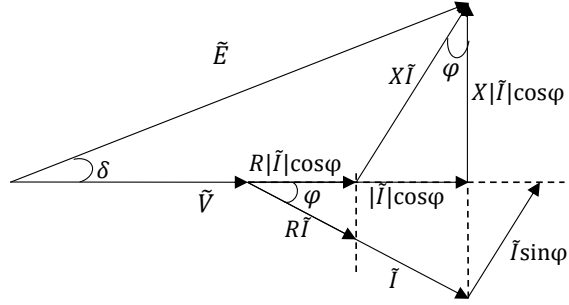}
        \caption{Electrical machine phasor diagram considering the machine resistance. Since the resistance is very low, its effect on the voltage is seen very small compared to that of reactance: the $R|\Tilde{I}|\cos\phi$ line is very small compared to the $X|\Tilde{I}|\cos\phi$ line.}
        \label{fig:phasor-1}
    \end{figure}

    To calculate the electrical torque produced by the machine, opposing the mechanical torque provided by the flywheel, we need to consider the electrical properties of the machine, hence to calculate the power it delivers. Let us assume that the terminal voltage $\Tilde{V}$ is a reference phasor. i.e. $\Tilde{V} = |\Tilde{V}|\angle 0$. Based on this, a phasor diagram is drawn in Fig. \ref{fig:phasor-1}. Here, it is assumed that the current $\Tilde{I}$ lags the voltage by an angle $\theta$ and the voltage $\Tilde{E}$ leads the terminal voltage $\Tilde{V}$ by an angle $\delta$. i.e. $\Tilde{I} = |\Tilde{I}|\angle\theta$ and $\Tilde{E} = |\Tilde{E}|\angle\delta$.
    
    Now, the current $\Tilde{I}$ can be written as,
    \begin{equation}
        \begin{split}
            \Tilde{I} &= \frac{\Tilde{V}-\Tilde{E}}{R+jX}\\
            %&= \frac{|\Tilde{V}|\angle 0-|\Tilde{E}|\angle\delta}{|Z|\angle\theta}\\
            &= \frac{|\Tilde{V}|}{|Z|}\angle(-\theta) - \frac{|\Tilde{E}|}{|Z|}\angle(\delta-\theta)\\
            % &= \left[ \frac{|\Tilde{V}|}{|Z|}\cos\theta - \frac{|\Tilde{E}|}{|Z|}\cos(\delta-\theta) \right] + i\left[ -\frac{|\Tilde{V}|}{|Z|}\sin\theta - \frac{|\Tilde{E}|}{|Z|}\sin(\delta-\theta) \right]
        \end{split}
        \label{eq:I}
    \end{equation}
    where $\theta = \arctan{(X/R)}$ is the angle associated with the machine parameters and $Z = \sqrt{R^2+X^2}$ is the internal impedance of the machine. From \eqref{eq:I}, we can write the equation for $|\Tilde{I}|^2$ as 
    \begin{equation}
    \begin{split}
        |\Tilde{I}|^2 = \frac{1}{|Z|^2}&\Big[ |\Tilde{V}|\cos\theta - |\Tilde{E}|\cos(\delta-\theta) \Big]^2 + \frac{1}{|Z|^2}\Big[ |\Tilde{V}|\sin\theta + |\Tilde{E}|\sin(\delta-\theta) \Big]^2\\
        % = \frac{1}{|Z|^2}&\Big[ |\Tilde{V}|^2\cos^2\theta + |\Tilde{E}|^2\cos^2(\delta-\theta) - 2|\Tilde{V}||\Tilde{E}|\cos\theta \cos(\delta-\theta)\\
        % & + |\Tilde{V}|^2\sin^2\theta + |\Tilde{E}|^2\sin^2(\delta-\theta) + 2|\Tilde{V}||\Tilde{E}|\sin\theta \sin(\delta-\theta) \Big]\\
        % = \frac{1}{|Z|^2}&\Big[ |\Tilde{V}|^2+|\Tilde{E}|^2 - 2|\Tilde{V}||\Tilde{E}|\cos[\theta+(\delta-\theta)] \Big]\\
         = \frac{1}{|Z|^2}&\Big[ |\Tilde{V}|^2+|\Tilde{E}|^2 - 2|\Tilde{V}||\Tilde{E}|\cos\delta \Big]
    \end{split}
        \label{eq:Isq}
    \end{equation}
    At the discharge speed $\omega_{\rm d}(t)$, the electrical torque reads \cite{kundur1994power}:
    \begin{equation}
    \begin{split}
        \overline{T_{\rm e}} &=\frac{P + R|\Tilde{I}|^2}{\overline{\omega_d}(t)}= \frac{|\Tilde{V}|}{|Z|\overline{\omega_d}(t)}\Big(|\Tilde{V}|\cos\theta-|\Tilde{E}|\cos(\theta-\delta)\Big) + \frac{|Z|}{\overline{\omega_d}(t)}\cos\theta|\Tilde{I}|^2\\
        %&=\frac{|\Tilde{V}|}{|Z|\overline{\omega_d}(t)}\Big(|\Tilde{V}|\cos\theta-|\Tilde{E}|\cos(\theta-\delta)\Big) + \frac{|Z|}{\overline{\omega_d}(t)}\cos\theta\frac{1}{|Z|^2}\Big( |\Tilde{V}|^2+|\Tilde{E}|^2 - 2|\Tilde{V}||\Tilde{E}|\cos\delta \Big)\\
        &=\frac{|\Tilde{V}|}{|Z|\overline{\omega_d}(t)}\Big(|\Tilde{V}|\cos\theta-|\Tilde{E}|\cos(\theta-\delta)\Big) + \frac{1}{|Z|\overline{\omega_d}(t)}\Big( |\Tilde{V}|^2+|\Tilde{E}|^2 - 2|\Tilde{V}||\Tilde{E}|\cos\delta \Big)\cos\theta
    \end{split}
    \end{equation}
    And as in per unit representation, $|\Tilde{V}|=1$ because our base voltage is the terminal voltage, the above equation becomes:
    \begin{equation}
        \overline{T_{\rm e}} = \frac{1}{|Z|\overline{\omega_{\rm d}}(t)}\Big(\cos\theta-|\Tilde{E}|\cos(\theta-\delta)\Big) + \frac{1}{|Z|\overline{\omega_{\rm d}}(t)}\Big(1+|\Tilde{E}|^2 - 2|\Tilde{E}|\cos\delta \Big)\cos\theta
        \label{eq:tele}
    \end{equation}
    
    Near an operating region where the machine is delivering some power to the grid, the generator voltage $\Tilde{E}$ does not change much, and because of the constant frequency, $|Z|$ is also constant; but $\overline{\omega_{\rm d}}(t)$ can be varying slightly due to friction. Hence, we can rewrite Eq.~(\ref{eq:tele}) as follows:
    
    \begin{equation}
        \overline{T_{\rm e}} = \frac{P_t}{\overline{\omega_{\rm d}}(t)}
        \label{eq:torqueK1}
    \end{equation}
    where $P_t$, the constant prefactor which can readily be identified from the electrical torque equation \eqref{eq:tele}, has the dimension of power and has value of the same around a normal operating region.
    
    \subsection{Linearization of torque equation}
    We know that the per-unit value of the angular speed of the machine during discharge $\overline{\omega_d}(t)<1$. If the flywheel speed goes too low, it cannot supply energy to the grid and the system must be disconnected. So, we may assume that during a normal discharge phase, the angular speed tends to 1, i.e. $\overline{\omega_d}(t)\longrightarrow 1$. To proceed with analytical calculations, we may define the function $g(t) = 1-\overline{\omega_d}(t)$ such that $g(t)\longrightarrow 0$; and after substitution in the torque equation (\ref{eq:torqueK1}), and expansion as a Maclaurin series we get:
    \begin{equation}
        \overline{T_{\rm e}} = P_{\rm t} \{1 + g(t) + [g(t)]^2 + [g(t)]^3 + \cdots\}
    \end{equation}
    Since $g(t)\longrightarrow 0$, we can neglect the second- and higher-order terms so that the per-unit electrical torque reads:
    \begin{equation}
        \overline{T_{\rm e}} \approx P_{\rm t}\{1 + g(t)\} = P_t\{1 + [1-\overline{\omega_{\rm d}}(t)]\} = P_t\{2-\overline{\omega_d}(t)\}
    \end{equation}
    which we differentiate to obtain 
    \begin{equation}
        \Dot{\overline{T_e}} = -P_t\{\Dot{\overline{\omega_d}}(t)\}
        \label{eq:TeDiffEq}
    \end{equation}

\begin{figure}[h]
        \centering
        \includegraphics[trim={6.5cm 12.5cm 5cm 2.95cm},clip,width = 0.8\textwidth]{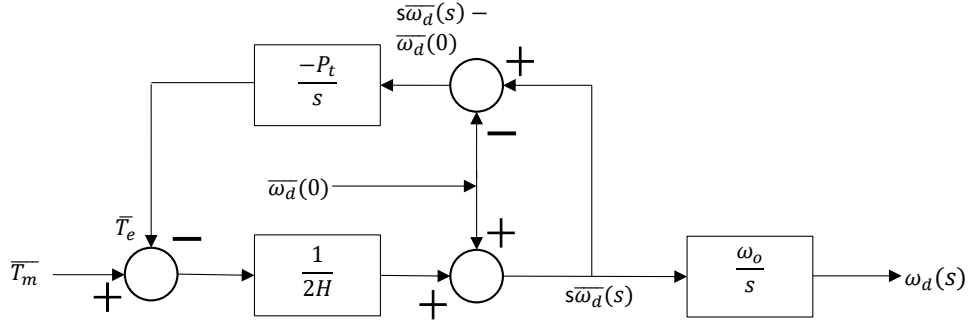}
        \caption{Block diagram of the electrical torque and speed relationship. The output of the flywheel (mechanical torque) is the input to this system whose output is the net speed. The term `$P_{\rm t}$' is the instantaneous active power delivered by the machine.}
        \label{fig:TFblockdiagram}
    \end{figure}

    The analysis of Eqs.~\eqref{eq:OmegaDiffEq} and \eqref{eq:TeDiffEq} may be done in the Laplace domain:
    \begin{eqnarray}
            s\overline{\omega_{\rm d}}(s) - \overline{\omega_{\rm d}}(0) &=& \frac{\overline{T_{\rm m}}-\overline{T_{\rm e}}}{2H}\\
            \overline{T_{\rm e}} &=& -P_{\rm t}\frac{s\overline{\omega_{\rm d}}(s) - \overline{\omega_{\rm d}}(0)}{s}
    \end{eqnarray}
    Note that on the onset of the discharge phase, there is initially no electrical power flow over the electrical machine, hence the initial electrical torque is zero: $\overline{T_e}(0) = 0$. The initial discharging speed of the flywheel $\overline{\omega_{\rm d}}(0)$ is the final charging speed of the flywheel: $\overline{\omega_{\rm d}}(0) = \overline{\omega_{\rm s}}(t_s)$. We can then easily draw a block-diagram representation of the torque-speed control system as shown in figure \ref{fig:TFblockdiagram}.
    
    There are now two ways to represent the input/output characteristics of a FESS. One is given in \eqref{eq:DischargeOmegaforSpeed}, where we can represent the instantaneous speed of the flywheel in terms of electrical speed lag, which could be derived from the electrical torque. The second one is the block diagram in Fig. \ref{fig:TFblockdiagram} from where we can derive the relation for discharge speed in terms of mechanical torque as follows:
    
    \begin{equation}
        \omega_{\rm d}(s) = \frac{sT_m + \overline{\omega}_{\rm d}(0)(2sH-P_{\rm t})}{(2sH-P_{\rm t})s} \omega_0
    \end{equation}
    where $\omega_0$ is the rated synchronous speed of the electrical machine and $\overline{\omega}_d(0)$ is the initial speed of the flywheel at the beginning of the discharge phase. 

    \subsection{Magnetic bearings and stabilization}
    Replacing the mechanical bearings of flywheels with magnetic bearings has given a huge advantage to FES systems by virtually completely eliminating the friction losses of medium- and long-term storage. The two technologies in use nowadays are permanent magnets bearings (PMB) and superconducting magnetic bearings (SMB) \cite{aljohani}. PMB use the properties of natural magnets for stabilization, while SMB use superconducting coils to generate a magnetic field. Further, the fine-tuned control of magnetic field based on gyroscopic measurements could be an effective method. The advantage of PMB is that they incur no power cost for their use, while their disadvantage lies in the fact that they cannot be controlled. SMBs however, do have a power cost to operate, but they can be controlled and hence can adapt to the real-time position of the flywheel for increased stabilization. Much of the stabilization for flywheels is necessary because of the vibrations from the motor, as well as angular momentum changes due to the rotation of the Earth. Some systems use a combination of SMB and PMB for optimal power cost and stabilization control \cite{komori}.

    \section{Discussion - Applications in electrical grids}
    FESS have three main ``sources'' of energy loss. The first two are conversion losses in the AC-to-AC converter and in the electrical machine, which occur both during the charging of electrical energy in the form of rotation kinetic energy and during the discharge phase due to electrical energy release. These losses depend entirely on the circuit design of the converter, and on the electrical machine design. The third source of energy loss is due to the stabilization process by the SMB. Note, however, that since any energy conversion technology entails losses, FESS remain a credible solution for storage. %Additionally, there is some frictional losses in the bearings during all the three phases of flywheel operation as mentioned in the earlier sections.
    
    As renewable energy sources become increasingly important in the electricity generation mix \cite{Lund2010}, particularly solar power and wind power, energy storage technologies are necessary to mitigate the irregular character of these energy resources \cite{hydro}. Several energy storage technologies are being developed and some of them are already in use today: pumped hydroelectric energy storage (PHES) is already used in several locations throughout the world, and  battery energy storage (BES) solutions, particularly lithium ion, have grown significantly over the past decade. Other technologies, such as super-capacitor storage and superconducting magnetic energy storage (SMES) are also being developed. In this playing field, FESS offer some actual advantages.
    
    To date, PHES is by far the most used technology for large scale energy storage, with a total installed capacity of 130 GW in 2016 \cite{hydro}. It has been in use since the mid-20$^{\rm th}$ century and offers the advantage of having a large capacity per installed unit (10-4000 MW). However, it has a quite extensive land use due to its low power density, which limits the technology to certain locations with the appropriate geographical factors.
    
    BES has become a more popular energy storage option recently, particularly with lithium ion batteries as its cost has strongly decreased and energy density strongly increased over the last decade \cite{comello}. One significant advantage of BES over PHES is that BES systems can be installed for individual households or buildings, taking up much less space. It also allows for a large network of many, small BES systems for a more flexible system. However, BES systems are very sensitive to temperature and work best for temperatures around 25$^{\circ}$C \cite{shim}. Furthermore, its most popular technology, lithium ion, is dependent on lithium mining, which is limited, relatively expensive, and has a significant negative environmental impact \cite{grosjean,chena}.
    
    While FESS boast several of the favorable characteristics of BES systems by virtue of being relatively small and dense energy storage technologies, they have the distinct advantage of not being sensitive to temperature, neither do they require the usage of rare metals with limited availability. Furthermore, FESS can deliver energy at much higher power then BES systems. We may thus reasonably expect that FESS will be a major energy storage option for electric grids over the next few decades \cite{joule,Wicki2017}.
    
    \begin{table}[h]
    \centering
    \caption{Characteristics for different electrical energy storage methods \cite{storage}}
    \begin{tabular}{llll}
    \hline\noalign{\smallskip}
    & BES (Li-ion) & FES  & PHES \\
    \noalign{\smallskip}\hline\noalign{\smallskip}
    Specific power [W/kg] & 150-315 & 400-1500 & 0.5-1.5\\
    Specific energy [Wh/kg] & 75-200 & 10-195 & 0.5-1.5\\
    Power Density [W/l] & 400-800 & 1000-2000 & 0.5-1.5\\
    Energy Density [Wh/l] & 200-500 & 80-270 & 0.5-1.5\\
    Lifetime [years] & 5-15 & $\sim$15 & 40-60+ \\
    Roundtrip Efficiency [$\%$] & 90-97 & 90-95 & 70-85\\
    \noalign{\smallskip}\hline
    \end{tabular}
    \end{table}

    \section{Concluding remarks}
    Flywheels have been in use for a long time in human history for different purposes. Due to recent technological developments, including the introduction of magnetic bearings and novel flywheel materials have made flywheel energy storage a competitive alternative to other energy storage solutions such as battery energy storage and pumped hydroelectric energy storage. In this work, we provided a detailed discussion of one of the main constraints in flywheel design: the mechanical stresses generated within the flywheel itself as it rotates. These latter depend on the geometry of the flywheel, the material used, as well as the angular velocity and angular acceleration of the flywheel. Equations (\ref{eq15}) in particular, may be used for quick yet informative assessment of composite flywheel design test cases with different materials \cite{Conteh2016}, provided that the appropriate boundary conditions are applied, and as long as the studied system is axisymmetric. We also analysed the design of the electrical systems that go together with the FESS. We have included a detailed mathematical model of the flywheel-electrical machine system which could be incorporated in electrical power system models for further study of stability. Furthermore, we considered a synchronous machine model in the present work but in case an induction motor would be considered, a slightly different control strategy would be required, though the basic principles of operation and analysis remain the same. For example, the AC-AC converter control should be adapted because an induction motor needs to be started at lower torque which could be achieved by high-voltage starting as opposed to low-frequency starting for a synchronous generator. Similarly, in the discharging phase, operation of an induction generator is different as its slip (change in speed with respect to synchronous speed) increases with discharge time.

% If in two-column mode, this environment will change to single-column format so that long equations can be displayed. 
%\begin{widetext}
%$$\mbox{put long equation here}$$
%\end{widetext}

% Figures should be put into the text as floats. 
% Use the graphics or graphicx packages (distributed with LaTeX2e).
% Put the label that you will use with \ref{} command in the braces of the \label{} command.
%
% \begin{figure}
% \end{figure}

% \begin{table}
% \end{tabular}
% \end{table}

\begin{acknowledgments}
P.-J. S. would like to thank the Skoltech Global Campus Program to allow for this research opportunity. S. G. would like to thank the Skoltech Energy Conversion Physics and Technology group for this research opportunity. H. O. acknowledges support by the Skoltech NGP Program (Skoltech-MIT joint project).
\end{acknowledgments}

\bibliography{RefBib}

\end{document}